\documentclass{elsart}
\usepackage{graphicx,amsmath,amssymb}
\journal{Physics Letters A}

\begin{document}

\newtheorem{Theorem}{Theorem}
\newtheorem{Prop}{Proposition}
\newtheorem{Coro}{Corollary}

\begin{frontmatter}

\title{Quantum operation, quantum Fourier transform and semi-definite programming}
\author[SKLITS]{Runyao Duan\corauthref{cor}\thanksref{thank}}
\ead{dry02@mails.tsinghua.edu.cn},
\author[SKLITS]{Zhengfeng Ji\thanksref{thank}}
\ead{jizhengfeng98@mails.tsinghua.edu.cn}
\author[SKLITS]{Yuan Feng\thanksref{thank}}
\ead{fengy99g@mails,tsinghua.edu.cn},
\author[SKLITS]{Mingsheng Ying\thanksref{thank}}
\ead{yingmsh@mail.tsinghua.edu.cn}
 \corauth[cor]{Corresponding
author.}
\address[SKLITS]{State Key Laboratory of Intelligent Technology and Systems,
Department of Computer Science and Technology Tsinghua University,
Beijing, China, 100084}

\thanks[thank]{This work was partly
supported by the National Foundation of Natural Sciences
of China (Grant No: 60273003).}

\begin{abstract}
We analyze a class of quantum operations based on a geometrical
representation of $d-$level quantum system (or qudit for short). A
sufficient and necessary condition of complete positivity,
expressed in terms of the quantum Fourier transform, is found for
this class of operations. A more general class of operations on
qudits is also considered and its completely positive condition is
reduced to the well-known semi-definite programming problem.
\end{abstract}


\end{frontmatter}

\section {Introduction}
Almost all quantum systems in the real world are open in the sense
that they suffer from unwanted interactions with the outside
world. The dynamics of an open quantum system is usually much more
complicated than that of a closed one. One of the most important
mathematical tools describing the dynamics of an open quantum
system is quantum operation, which has been systematically studied
after Kraus's seminal work \cite{KRAUS83}. Comparing with other
mathematical formalisms coping with open quantum systems, such as
master equation, Langevin equation and stochastic differential
equation, quantum operations are especially appropriate for
depicting discrete state change in discrete time. Recently, the
rapid development of quantum information processing technology
revives a wide interest on quantum operations due to the fact that
quantum information processing systems, for example, quantum
computers, suffer from outside noises inevitablly, and that mainly
a discrete-time evolution is concerned in these systems. Indeed,
quantum operation plays an important role in many active fields
such as quantum computation, quantum information, quantum
error-correcting codes and quantum fault-tolerant computation
\cite{Got98,CS96,Sho95} (for an excellent exposition,
see \cite{MANILC} Chapter 8).

There are several equivalent ways to introduce the notion of
quantum operation; one of which is given in terms of completely
positive (often abbreviated as CP) mappings. Let $\mathcal{H}$
denote the Hilbert space of the principal quantum system. A linear
mapping $\kappa$ on $\mathcal{H}$ is positive if it always sends a
positive operator to another positive one; and $\kappa$ is
completely positive if, furthermore, the mapping $I \otimes
\kappa$ is also positive where $I$ is the identity mapping acting
on an arbitrary ancillary system. Complete positivity is a natural
requirement of the real physical world without which the state of
the composite system may be invalid after operations on its
subsystem. Thus, a quantum operation is defined as a completely
positive and trace-preserving linear mapping on the state space
$\mathcal{H}$. Some representations of the CP mapping have been
presented in \cite{KRAUS83}, also in\cite{Choi}, and an important
one is the operator sum representation:
\begin{equation}
\mathcal{E}(\rho)=\displaystyle\sum_{k=1}^{N}A_{k}\rho
A_{k}^{\dagger},
\end{equation}
where $A_{k}^{\dagger}$ is the conjugate transpose of $A_{k}$.
To guarantee the trace-preserving property, the
following completeness condition
\begin{equation}
\displaystyle\sum_{k=1}^{N}A_{k}^{\dagger}A_{k}=I
\end{equation}
is also required. For some other descriptions of quantum
operations, we refer to \cite{MANILC}. As trace-preserving
quantum operation is a fundamental mathematical description of
quantum channel, a thorough study on quantum operation will help
us to understand the limits or capabilities of quantum information
processing.

The main goal of this paper is to find the necessary and
sufficient conditions for certain kinds of linear mappings to be
quantum operations. To motivate our problem, let us consider the
trace-preserving quantum operation $\kappa$ on a two-level quantum
system (or qubit for short). We employ the very useful geometric
tool of Bloch sphere to represent a qubit. Then every qubit can be
depicted as a vector in it. Recall that the general form of the
affine mapping on the Bloch sphere is
\begin{equation}
\vec{r}=(r_x,r_y,r_z)\mapsto
\vec{r}'=M(r_x,r_y,r_z)+(c_x,c_y,c_z),
\end{equation}
where $\vec{r}$ and $\vec{r}'$ are respectively the original and
the image vectors under the mapping, $M$ is a $3\times3$ real
matrix and $\vec{c}=(c_x,c_y,c_z)$ the displacement vector. It is
easy to see that the effect of a quantum operation on qubits is
just an affine mapping on the corresponding vectors. Conversely, a
question naturally arises: whether or not every affine mapping on
the Bloch sphere has a corresponding quantum operation? The answer
to this question is unfortunately no. In fact, there are very
simple affine maps that cannot be images of quantum operations
\cite{AP99,AH98}. A sufficient and necessary condition of
the special case without any displacement has been found by
several research groups \cite{AP99,BIDMJ99,CM01} and
some special cases of $(c_x,c_y,c_z)$ are also considered. In
\cite{MSE02}, M. B. Ruskai, S. Szarek and E. Werner have
completely solved the general problem for the qubit case. More
precisely, given the transition matrix $M$ and displacement $\vec{c}$,
the sufficient and necessary condition of when such a mapping is
completely positive was presented there.

A qubit is the simplest quantum system which has been studied
thoroughly. However, quantum systems of higher dimension also concern us
in quantum information processing and are relatively less studied.
This observation leads us to consider higher dimensional generalization
of the above question. To the authors' best knowledge, the higher
dimensional version of this question is still open. In the present paper,
we will carefully examine the affine mappings in higher dimensional quantum systems,
and give a partial answer to this question. This will certainly
give some new insights to the field of quantum information
processing and bring a useful mathematical tool for the solutions
of some problems in this field.

The paper is organized as follows. In Section $2$, we extend the
vector representation from qubit to qudit case. This enables us to
give a $d$-dimensional generalization of affine mappings. Some
simple properties of such geometric representation are presented.
In Section $3$, we focus on a special class of higher dimensional
affine mappings where the transition matrix is diagonalized and
the displacement vector vanishes. A sufficient and necessary
condition of when an mapping in this class is a quantum operation
is found. With this condition, one can easily check whether a
given mapping in the special case is an image of a quantum
operation or not. Some well known results, for example, the
universal NOT gate with optimal fidelity in the high dimensional
state space, can be recovered. A more general case involving
multiple qudits is also discussed. The most interesting thing is
that this condition is deeply related to the quantum Fourier
transform (QFT for short), and such a surprising connection is not
easy to observe if we only deal with the case of qubits. In
Section $4$, we further consider higher dimensional affine
mappings with diagonal transition matrix in which the displacement
vector presents. They are of the most common affine mappings in
practical use. The problem of determining when such an affine
mapping is a quantum operation is then reduced to the
semi-positive definite programming problem. We argue that this
problem has no simple solution in the general case. However, we
are still able to give a simple sufficient condition of it.
Finally, we draw a brief conclusion in Section $5$.

\section{A geometric representation of qudits}
It is well known that for qubit system, a wonderful geometric
representation using Bloch sphere exists. To be specific, the
density matrix $\rho$ of a qubit can be written as
$$\rho=\frac{I+\vec{r}\cdot\vec{\sigma}}{2},$$
where $\vec{r}=(r_x,r_y,r_z)$ is a $3$-dimensional real vector
with Euclidean norm $||\vec{r}||\leq1$. $\vec{\sigma}=(X,Y,Z)$ is
a formal vector of Pauli matrices and $\vec{r}\cdot
\vec{\sigma}=r_xX+r_yY+r_zZ$ where $X, Y, Z$ are Pauli matrices,
$$X=\begin{pmatrix}
0 & 1\\
1 & 0 \\
\end{pmatrix},Y=\begin{pmatrix}
0 & -i \\
i & 0
\end{pmatrix},Z=\begin{pmatrix}
1 & 0 \\
0 & -1
\end{pmatrix}.$$
It is notable that $||\vec{r}||= 1$ if and only if $\rho$ is a
pure state, and $\rho$ is a mixed state whenever $||\vec{r}||<1$.
Given a vector $\vec{r}$, we can completely determine the
corresponding quantum state $\rho$ and vise versa. Furthermore,
the dynamics of $\rho$ can be visualized through the affine
operation on $\vec{r}$. The key of a qubit possessing the Bloch
sphere vector representation is there exist a set of matrices
$\{X,Y,Z\}$ which, together with the identity matrix $I$, forms a
basis of the $2 \times 2$ matrix space.

To extend such a representation to a higher dimensional state
space, we choose a set of generalized Pauli matrices
\cite{Got98} as the basis. Suppose we have an orthonormal basis \{$|k\rangle$\}
of a qudit space $\mathcal{H}$, if we define $X$ as
$X|k\rangle=|k+1\rangle$, where the addition is modulo $d$, and
$Z$ as $Z|k\rangle=\omega^{k}|k\rangle$, where
$\omega=exp(-\frac{2\pi i}{d})$ is the $d$th unity root, then the
whole set of the general Pauli matrices in qudit system can be
expressed as $\{\sigma_{p,q}=X^{p}Z^{q}:p,q=0,\cdots,d-1\}$. For a
detailed presentation of the properties the generalized Pauli
matrices enjoy, we refer to \cite{Got98}. Here we only list some
of them related to our results (for simplicity, we define
$X_i=X^{i},Z_i=Z^{i}$):

\begin{enumerate}
\item (Multiplication relations)
$$\sigma_{i,j}\sigma_{i',j'}=\omega^{i'j}\sigma_{(i+ i'),( j+ j')}.$$

\item (Hermitian relations)
$$X_i^\dagger=X_{d-i},\ Z_i^\dagger=Z_{d-i},$$
$$\sigma_{i,j}^\dagger = \omega^{ij}\sigma_{(d-i),(d-j)}.$$

\item (Commutation relations)
$$\sigma_{i, j}\sigma_{i', j'}=\omega^{(i' j-i j')}\sigma_{i', j'}\sigma_{i, j}.$$

\item (Orthogonal relations) Suppose the inner product of $d\times
d$ linear space is defined as $(A,B) = tr(A^{\dagger} B)$, then
$$(\sigma_{i,j},\sigma_{i', j'}) = d \delta_{i,i'}\delta_{j,j'},$$
where $\delta$ is the Kronecker function.
$$\delta_{i,j}=\left\{
\begin{array}{cc}
0, & if \ \ i\neq j \\
1, & if \ \ i=j
\end{array}
\right. .$$

\item The Pauli matrices $\{\sigma_{i,j}:\ i,j=0,1,\cdots,d-1\}$
form an orthogonal basis of $d\times d$ linear space $Q$. That is,
for any $A\in Q$,
$$A=\frac{\sum_{i,j}
tr(\sigma_{ij}^{\dagger}A)\cdot\sigma_{i,j}}{d}.$$
\end{enumerate}

The next proposition shows that on a higher dimensional space the
matrix composed of all eigenvectors of $X$ corresponds exactly to
the Fourier transform $F$.

\begin{Prop}
Suppose $F$ is the quantum Fourier transform on $d$-dimensional Hilbert
state space $\mathcal{H}$, that is
$$F|j\rangle = \frac{1}{\sqrt{d}}\sum\limits_{k=0}^{d-1} e^{2\pi ijk/d}|k\rangle,$$
then $X_k=F^{\dagger}Z_kF$, where $\{|j\rangle :
j=0,1,2,\cdots,d-1\}$ is an orthonormal basis of $\mathcal{H}$.
\end{Prop}
{\bf Proof.} Simple, details omitted.\qed

Using this set of generalized Pauli matrices in high dimensional
space, we can decompose a density matrix $\rho$ of a qudit in the
principal Hilbert space $\mathcal{H}$ as follows:
 $$\rho=\frac{\vec{r}\cdot\vec{\sigma}}{d},$$
where
$\vec{r}=(tr(\sigma_{00}^{\dagger}\rho),tr(\sigma_{01}^{\dagger}\rho),
\cdots,tr(\sigma_{d-1,d-1}^{\dagger}\rho))$ is a $d^2-$dimensional
vector and $\vec{\sigma}$ is the formal vector such that
$\vec{\sigma}=(\sigma_{0,0},\sigma_{0,1},\cdots,\sigma_{d-1,d-1})$,
the formal inner product is then defined as
$$\vec{r}\cdot\vec{\sigma}=\sum_{p,q=0}^{d-1} r_{p,q}\sigma_{p,q}.$$
The representation here is compatible with the Bloch vector notation
since we can simply add $I$ and $1$ to the Pauli matrices $\vec{\sigma}$ and $\vec{r}$ respectively.
Moreover, some properties of Bloch vector notation preserve in the new
representation. For example, $||\vec{r}||=\sqrt{d}$ when
$\rho$ is a pure state; $||\vec{r}||<\sqrt{d}$ when $\rho$ is a
mixed state, while $||\vec{r}||=1$ if and only if $\rho$ is the
maximally mixed state $I/d$. It is also useful to point out that
to make $\rho$ a legal quantum state, we always have
$r_{0,0}=tr(\sigma_{0,0}^{\dagger}\rho)=tr(\rho)=1$.

Suppose quantum operation $\mathcal{E}$ on $\mathcal{H}$ maps
$\rho$ to $\rho'$ with the representation
\begin{equation}
\rho'=\frac{\vec{r}'\cdot\vec{\sigma}}{d},
\end{equation}
a simple computation tells us there exists a $d^2\times d^2$
dimensional matrix $M$ such that
$$\vec{r}'= M\vec{r}.$$ In other words, every quantum operation leads to
an affine mapping with the form
\begin{equation}
\vec{r} \mapsto M\vec{r}. \label{general}
\end{equation}

Then the question raised in the introduction can be more
explicitly stated as follows: is every affine mapping with the
form (\ref{general}) a quantum operation on state space
$\mathcal{H}$? In general, the answer to this question is
certainly no. Our purpose of this paper is then to characterize
the affine mapping that corresponds to some quantum operation.

Before we go deep into this question, let us first examine some
properties that transition matrix $M$ and vector $\vec{r}$ must
satisfy. We say a quantum operation $\mathcal{E}$ is unital if
$\mathcal{E}(I)=I$. Otherwise we call it is non-unital. Suppose
transition matrix $M$ corresponds to a trace-preserving quantum
operation $\mathcal{E}$, then $M$ has the block form
 $$M=\begin{pmatrix}
1 & 0\\
c & M' \\
\end{pmatrix},$$ where $c$ is a $d^2-1$ dimensional
vector(we often call it as displacement vector), $M'$ is a
$(d^2-1)\times (d^2-1)$ matrix. Furthermore, if $\mathcal{E}$ is
unital, we have $c=0$ (vector). For $\rho$ is a positive operator
with trace one, we get $\vec{r}\cdot\vec{\sigma}=
(\vec{r}\cdot\vec{\sigma})^{\dagger}$ and $r_{00}=1$, or
equivalently, $$r_{p,q}^{\ast}=\omega^{-pq}r_{-p,-q},r_{00}=1,$$
where all the arithmetical operations of the indices are modulo
$d$, so $r_{-p,-q}=r_{d-p,d-q}$. The above equations can be
considered as a general constraints on the vector $\vec{r}$. With
this condition, we can use $d^2-1$ real numbers to represent a
qudit state $\rho$.

Denote the set of vectors satisfying the above equations as
$V_{d^2}$. A simple calculation shows that $V_{d^2}$ is close
under addition and multiplication by a real number; that is,
$\vec{r_1}+\vec{r_2}\in V_{d^2}$ and $t\vec{r_1}\in V_{d^2}$
provided that $\vec{r_1},\vec{r_2}\in V_{d^2}$ and $t$ is a real
number. Notice that when $t$ is a complex number, $t\vec{r_1}\in
V_{d^2}$ does not necessarily hold. Suppose A is a $d^2\times d^2$
matrix which is a linear mapping on $V_{d^2}$, then $A\vec{r}\in
V_{d^2}$ for all $\vec{r}\in V_{d^2}$ if and only if
$(A\vec{r}\cdot \vec{\sigma})^{\dagger}=A\vec{r}\cdot
\vec{\sigma}$ for any $\vec{r}\in V_{d^2}$. Let us further
consider a special but very important case where $A$ is a diagonal
matrix
$$A=diag(\lambda_{0,0},\lambda_{0,1},\cdots,\lambda_{d-1,d-1}),$$ or simply we write
$A=diag(\vec{\lambda}).$
 Then the condition turns out to
be $\lambda_{-p,-q}=\lambda_{p,q}^{\ast}$. So if a quantum
operation maps $\rho$ with the Bloch vector $\vec{r}$ to
$\mathcal{E}(\rho)$ with the Bloch vector $M\vec{r}$, where
$$M=diag(\lambda_{0,0},\lambda_{0,1},\cdots,\lambda_{d-1,d-1})=diag(\vec{\lambda})$$ is diagonal,
then the following conditions must be satisfied:
$$r_{p,q}=\omega^{pq}r_{-p,-q}^{\ast},\lambda_{-p,-q}=\lambda_{p,q}^{\ast}.$$
For the case of qubits, i.e., $d=2$, we have the pauli matrices
 $$\sigma_{0,0}=I,\ \sigma_{0,1}=Z,\ \sigma_{1,0}=X,\
 \sigma_{1,1}=XZ$$ and $\omega=-1$,
then the above conditions about $\vec{r}$ may be rewritten as
$$r_{0,0}=1,\ r_{0,1}=r_{0,1}^{\ast},\ r_{1,0}=r_{1,0}^{\ast},r_{1,1}=-r_{1,1}^{\ast};$$
that is, $r_{0,0}=1,r_{0,1},r_{1,0}$ are real numbers and
$r_{1,1}$ is pure imaginary number. The conditions about $M$ can
be rewritten as
$$\lambda_{0,0}=1,\lambda_{0,1}=\lambda_{0,1}^{\ast},\lambda_{1,0}=\lambda_{1,0}^{\ast},\
\lambda_{1,1}=\lambda_{1,1}^{\ast},$$ all the four numbers
$r_{0,0}, r_{0,1}, r_{1,0}$ and $r_{1,1}$ are real. Moreover, by
introducing $Y=iXZ$, we can also make $r_{1,1}$ a real
number. Then the qubit state $\rho$ and real vector
$\vec{r}=(r_{0,0},r_{0,1},r_{1,0},r_{1,1})$ are one to one
corresponding to each other when $|\vec{r}|\leq 2$, and the
scalars of the four axes are all real numbers.

\section{Affine mappings with diagonal transition matrix}
In this section, we consider the special affine mappings with
diagonal transition matrices $M$ and no displacement vectors. The
effect of these mappings is just a scalar multiplication by
$\lambda_{p,q}$ in the direction $\sigma_{p,q}$, or
\begin{equation}
\vec{r} \mapsto M\vec{r}, \label{diag}
\end{equation}
where $M=diag(\vec{\lambda})$ is a diagonal matrix. we can also
write
$$\mathcal{E}(\sigma_{p,q})=\lambda_{p,q}\sigma_{p,q},\ \ \ p,q=0,\cdots, d-1.$$
It is well known that when qubits are considered, with a suitably
chosen axis, many interesting quantum operations have this simple
form in the Bloch sphere. The typical examples are bit flip
channel and phase damping \cite{MANILC}.

This section is devoted to find the necessary and sufficient
condition of such affine mapping to be a quantum operation's
image. As mentioned before, a key feature of a quantum operation
is its complete positivity. To determine the complete positivity
of $\mathcal{E}$, we need the following theorem attributed to
Kraus \cite{KRAUS83}:

\begin{Theorem} Suppose $\mathcal{E}$ is a linear operation on Hilbert space
$\mathcal{H}$, $\mathcal{R}$ is an ancillary Hilbert space with
the same dimensionality with $\mathcal{H}$. Let $\{|k\rangle\}$ be
an orthonormal basis for $\mathcal{H}$ and $\mathcal{R}$. Then
$\mathcal{E}$ is CP mapping if and only if the operator
$$(I\otimes \mathcal{E})(|\alpha\rangle\langle\alpha|)$$ is
positive, where
$$|\alpha\rangle=\displaystyle\sum_{k}|k\rangle|k\rangle /
\sqrt{d}$$ is the maximal entangled state of the composite system.
\end{Theorem}

Using Kraus's theorem, to decide whether a given linear mapping is
a CP one, we need only examine the positivity of eigenvalues of
the operator $(I\otimes
\mathcal{E})(|\alpha\rangle\langle\alpha|)$.

We now are able to present one of the
main results of this paper, namely, a CP condition for an affine
mapping $\mathcal{E}$ with a diagonal transition matrix and no
displacement vector. To one's surprise, this condition is
essentially connected to the quantum Fourier transform $F$,
which, as we all known, plays a crucial role in quantum computation
(for example, quantum Fourier transform is a key step for the
Shor's Factoring algorithm and Discrete Logarithm
Problem\cite{Sho95}). To be specific, we have the following
theorem.

\begin{Theorem}\label{thm:qftne}
The linear mapping defined in (\ref{diag}) is a CP one if and only
if
$$(F\otimes F^{\dagger}){\vec{\lambda}}\geq0.$$
where
$\vec{\lambda}=(\lambda_{0,0},\lambda_{0,1},\cdots,\lambda_{d-1,d-1}).$
\end{Theorem}
{\bf Proof.} Using the Kraus's theorem, we need to calculate
$(I\otimes \mathcal{E})(|\alpha\rangle\langle\alpha|)$. So first
we expand $|k\rangle \langle l|$ as the linear combinations of
$\{\sigma_{i,j}\}$. By the orthogonality of $\sigma_{i,j}$, we get
$$|k\rangle \langle
l|=\displaystyle\sum_{m,n}\alpha_{m,n}\sigma_{m,n},$$ where
$\alpha_{m,n}=tr(\sigma_{m,n}^{\dagger}|k\rangle\langle
l|)/d=\omega^{-nl}\delta_{m+l,k}/d.$ So
\begin{eqnarray}
\begin{array}{rl}
&(I\otimes \mathcal{E})(|\alpha\rangle\langle\alpha|)\\
\\
=&\displaystyle\sum_{k,l}|k\rangle \langle
l|\mathcal{E}(|k\rangle \langle l|)/d\\
=&\displaystyle\sum_{l,m,n,q}\omega^{-(q+n)l}\sigma_{m,q}\otimes\sigma_{m,n}\lambda_{m,n}/d^3\\
=&\displaystyle\sum_{m,n}\lambda_{m,n}\sigma_{m,-n}\otimes\sigma_{m,n}/d^2.\label{able}
\end{array}
\end{eqnarray}
The crucial point of the problem is that the
set$$\{\sigma_{m,-n}\otimes \sigma_{m,n}\}$$ form an Abelian group
of order $d^2$, which means that this set is closed under the
operation of matrix multiplication, matrix inversion, and its
elements commute with each other. We prove commutation relation
as follows:
\begin{eqnarray}
\begin{array}{rl}
&(\sigma_{m,-n}\otimes \sigma_{m,n})(\sigma_{p,-q}\otimes\sigma_{p,q})\\
\\
=&(\sigma_{m,-n}\sigma_{p,-q})\otimes(\sigma_{m,n}\sigma_{p,q}) \\
\\
=&(\omega^{-np+qm}\sigma_{p,-q}\sigma_{m,-n})\otimes
(\omega^{np-qm}\sigma_{p,q}\sigma_{m,n}) \\
\\
=&(\sigma_{p,-q}\otimes\sigma_{p,q})(\sigma_{m,-n}\otimes\sigma_{m,n}).
\end{array}
\end{eqnarray}
Thus, we can simultaneously diagonalize the set of matrices
$\{\sigma_{p,-q}\otimes \sigma_{p,q}\}$ as $$\sigma_{m,-n}\otimes
\sigma_{m,n}=UD(m,n)U^{\dagger},$$ where $D(m,n)$ is a diagonal
matrix, and $U$ a unitary matrix. It must be noted that the matrix
$U$ can be chosen as the same for all the matrices in the set.
Then$$(I\otimes
\mathcal{E})(|\alpha\rangle\langle\alpha|)=\frac{1}{d^2}\displaystyle\sum_{m,n}U\lambda_{m,n}D(m,n)U^{\dagger}.$$
The diagonal entries give the desired eigenvalues of $(I\otimes
\mathcal{E})(|\alpha\rangle \langle\alpha|)$. Define
$$|\Phi_{s,t}\rangle=\frac{1}{\sqrt{d}}\displaystyle\sum_{k}|k\rangle
\sigma_{t,s}|k\rangle,$$ one can easily check
$$(\sigma_{m,-n}\otimes
\sigma_{m,n})|\Phi_{s,t}\rangle=\omega^{nt-sm}|\Phi_{s,t}\rangle.$$
So by (\ref{able}) we have
$$(I\otimes
\mathcal{E})(|\alpha\rangle\langle\alpha|)|\Phi_{s,t}\rangle=\mu_{s,t}|\Phi_{s,t}\rangle,$$
where
\begin{eqnarray}
\begin{array}{rl}
\vec\mu_{st}&=1/d^2\displaystyle\sum_{m,n}\omega^{nt-sm}\lambda_{m,n}\\
&=1/d\displaystyle\sum_{m,n}(F)_{s,m}(F^{\dagger})_{t,n}\lambda_{m,n}\\
&=1/d[(F\otimes F^{\dagger})\vec{\lambda}]_{s,t},
\end{array}
\end{eqnarray}
or in matrix representation,
$$\vec\mu=\frac{1}{d}(F\otimes F^{\dagger})\vec{\lambda}.$$ For $(I\otimes
\mathcal{E})(|\alpha\rangle\langle\alpha|)$ is a Hermitian matrix,
it is positive if and only if all its eigenvalues are
non-negative. This leads to the relation $$(F\otimes
F^{\dagger})\vec{\lambda}\geq 0.$$ With this we complete the proof
of the theorem.\qed

There are many ways to gain the result in the
theorem and some of them are simpler. But the method given
here is a coherent one in the sense that when the displacement vector
does not vanish it still works.

Now we consider further the case where $N$ qudits are involved.
Suppose the generalized Pauli matrices on $d$-dimensional Hilbert
space $\mathcal{H}$ are $\{\sigma_{p,q}:p,q=0,\cdots,d-1\}$. Use
this set Pauli matrices, we can obtain an orthogonal basis of
operators acting on $\mathcal{H}^{\otimes N}$. For the sake of
convenience, we denote $i=(i_1,i_2,...,i_N)$ and
$j=(j_1,j_2,...,j_N)$, then we define the orthogonal basis as
$\pi_{i,j}=\sigma_{i_1,j_1}\otimes\sigma_{i_2,j_2}\cdots,\otimes\sigma_{i_N,j_N}$
,where $i_1,j_1,\cdots,i_N,j_N$ are range from $0$ to $d-1$. Then
the following theorem is a straightforward generalization of
Theorem \ref{thm:qftne}.

\begin{Theorem} Let $\mathcal{H}$ be a $d$-dimensional Hilbert space,
and let $\mathcal{E}$ be a linear mapping on Hilbert space
$\mathcal{H}^{\otimes N}$ such that
$$\mathcal{E}(\pi_{i,j})=\lambda_{i,j}\pi_{i,j}.$$ Then
$\mathcal{E}$ is CP if and only if
$$(F\otimes F^{\dagger})^{\otimes N}{\vec{\lambda}}\geq 0.$$
\end{Theorem}

{\bf Proof.} Same as the previous theorem. We omit the details
here.

We should point out here that the quantum Fourier
transform $F$ in the above theorem is defined on $\mathcal{H}$.
There is another Fourier transformation $F'$ which is defined on
$Nd$-dimensional state space $\mathcal{H}^{\otimes N}$. $F'$ is
not equal to $F^{\otimes N}$ in general. This enables us to obtain
two different conditions. However, this two conditions are not
related since they describe different types of affine mappings
which are directly related to the choice of basis. One basis is
general Pauli matrices and the other is tensor products of general
Pauli matrices.

As a conclusion of this section, we show
some simple applications of the above two theorems. First, let us
come back to the case of qubits. Notice that in this case, quantum
Fourier transform $F$ is just the $Hadamard$ gate
$$H=(X+Z)/ \sqrt{2}=\frac{1}{\sqrt{2}}\begin{pmatrix}
1 &1 \\
1 & -1
\end{pmatrix}.$$ For $H^{\dagger}=H$, the CP condition given
in Theorem 3.3 can be rewritten in a more brief form: $$H^{\otimes
2N}\vec{\lambda}\geq 0.$$ When only one qubit involves, the
condition deduces to $H^{\otimes 2}\vec{\lambda}$, or more
explicitly(noticing that $\lambda_{0,0}=1$)
 $$1+ \lambda_{0,1}+ \lambda_{1,0} +\lambda_{1,1}\geq 0,$$
 $$1- \lambda_{0,1}+ \lambda_{1,0}-\lambda_{1,1}\geq 0,$$
 $$1+ \lambda_{0,1}- \lambda_{1,0}-\lambda_{1,1}\geq 0,$$
 $$1- \lambda_{0,1}- \lambda_{1,0}+\lambda_{1,1}\geq 0,$$
which is similar as the equation (12) in \cite{MSE02}.

Since $(F\otimes F^{\dagger})^{\otimes N}{\vec{\lambda}}$ has an entry
$\sum \lambda_{i,j}/d^N$, a necessary condition for $\mathcal{E}$
to be a CP mapping is that $\sum
\lambda_{i,j}\geq 0$. An interesting special case here is when all
entries in $\vec{\lambda}$ have some same real
value $p$ except $\lambda_{0,0}=1$. This is in
fact a generalized depolarizing channel. In this case, we have two
different eigenvalues: $1+(d^2-1)p$ and $1-p$. The CP condition
then becomes
$$-1/(d^2-1)\leq p \leq 1.$$ This result is well known and here it
comes out to be a direct corollary of our above theorem.

What we would like to point out here is that when $c$ is negative
, $-1/(d^2-1)$ is the limit value we can achieve in this
operation. It is also easy to see that this value decreases with
$d$ increasing. This means that when dimensionality increases, the
state space becomes more complicate, and the constraints between
the different axes are more rigor.

Let us consider further the case where
$p=-1/(d^2-1)$. The pure state $\rho=|\psi\rangle \langle \psi|$
will be mapped to $\mathcal{E}(\rho)=(dI-\rho)/(d^2-1)$. If we
treat the operation as an unprecise universal NOT operation, then
the fidelity will be
\begin{equation}
\langle \psi^{\bot}
|\mathcal{E}(\rho)|\psi^{\bot}\rangle=\frac{d}{d^2-1},
\end{equation}
which has been proved optimal in the case of $d=2$, see
\cite{VMR99,JF01}.

Recall from Section 2 that we have the constraints
$\lambda_{p,q}=\lambda_{-p,-q}^{\ast}$. So these operations are
not completely independent in different axes, and there are
conjugate axes such as $\sigma_{p,q}$ and $\sigma_{-p,-q}$. Thus,
the scalars of a CP operation in this two directions must be
conjugated with each
other. In the qubit case, since $\sigma_{i,j}$ are all
self-conjugated, the entries of $\vec{\lambda}$ are all real
numbers, and the operation is independent for different axes.

\section{A more general class of affine mappings}

What concerns us in the previous section are affine mappings without
displacement vectors. In this section, we consider a more general
affine mapping of the form:
\begin{equation}
\vec{r}\mapsto M\vec{r}\label{addc},
\end{equation}
where $$M=diag(\vec{\lambda})+\begin{pmatrix}
0 & 0\\
c & 0 \\
\end{pmatrix}
.$$ We denote the first column of $\begin{pmatrix}
0 & 0\\
c & 0 \\
\end{pmatrix}$ as $\vec{c}$. Intuitively, we call $\vec{c}$ as
displacement vector. Such kind of affine mappings correspond the
most useful non-unital quantum operations in practice. The problem
in which we are interested remains when such an affine mapping is
an image of a quantum operation. The following theorem is the
another main results in our paper which establishes a connection
between this problem and the semi-positive definite programming.

\begin{Theorem}
The linear mapping defined in (\ref{addc}) is a CP one if and only
if
$$ I\otimes \vec{c}\cdot\vec{\sigma}+\displaystyle\sum_{p,q}\lambda_{p,q}\sigma_{p,-q}\otimes\sigma_{p,q}\geq 0.$$
\end{Theorem}

{\bf Proof.} From
$$\mathcal{E}(\sigma_{0,0})=\lambda_{0,0}\sigma_{0,0}+\vec{c}\cdot\vec{\sigma},$$we get
$$\mathcal{E}(\sigma_{p,q})=\lambda_{p,q}\sigma_{p,q},\ \ (p,q)\neq(0,0)$$ and
$$(I\otimes \mathcal{E})(|\alpha\rangle \langle\alpha|)=\frac{1}{d^2}[I\otimes\vec{c}\cdot
\vec{\sigma}+\displaystyle\sum_{p,q}\lambda_{p,q}\sigma_{p,-q}\otimes
\sigma_{p,q}].$$ Then by the Kraus's Theorem, the claim in the
theorem holds, and we complete the proof.\qed

Since the set of matrices $\{\sigma_{p,-q}\otimes \sigma_{p,q}\}$
can be simultaneously diagonalized, we can rewrite the condition
in the above theorem as
$$U^{\dagger}(I\otimes \vec{c}\cdot\vec{\sigma})U+ D\geq 0,$$
where $D=diag((F\otimes F^{\dagger})\vec{\lambda})/d.$

There are many basic problems in quantum information fields which
can be reduced to the most general version of the semi-positive
definite programming problem; for examples, quantum state
discrimination \cite{CH00,SUN02}, state estimation
\cite{KH89}, and quantum pattern recognition \cite{KABDM02}. The
above theorem provides us with an additional example of such
problems. Indeed, given $\vec{c}$. To keep $\mathcal{E}$ as a CP
mapping, we need to determine the range of the diagonal matrix
$D$. Then through $D$ we can immediately get the range of
$\vec{\lambda}$, for
$$\vec{\lambda}=d(F^{\dagger}\otimes F)\vec{D},$$where $\vec{D}$
denotes the vector formed by the diagonal entries of $D$.
Furthermore, the problem of determining the range of a diagonal
matrix $D$ to keep $A+D\geq 0$ for a Hermitian matrix $A$ is
exactly a semi-positive definite
programming problem.

As is well known, there has not yet a general analytic solution
for the problem of semi-positive definite programming. We then
cannot get a sufficient and necessary condition in an explicit
form under which the linear mappings (\ref{addc}) are completely positive
for the most general case. Here we propose instead a sufficient
condition:
\begin{Coro}
A sufficient condition for the linear mapping defined in
(\ref{addc}) to be a CP one is
\begin{equation}
||\vec{c}||\leq \mu_{\rm min} ,\label{suff}
\end{equation}
where $$\mu_{\rm min}={\rm min}_{s,t}[(F\otimes
F^{\dagger})\vec{\lambda}]_{s,t}/d,$$ and
$$||\vec{c}||=\sqrt{\sum_{p,q}|c_{p,q}|^2}$$ is the norm of the
displacement vector.
\end{Coro}

The intuitive meaning of the conclusion is that
the affine mapping (\ref{addc}) will be an image of a quantum
operation if the norm of the displacement vector $\vec{c}$ is
small enough. Of course, here $\vec{c}$ should satisfy the
equation $\vec{c}\cdot \vec{\sigma}=(\vec{c}\cdot
\vec{\sigma})^{\dagger}$.

{\bf Proof.} Let $\vec{c}\cdot\vec{\sigma}=VD_cV^{\dagger}$ be the
spectral decomposition of $\vec{c}\cdot\vec{\sigma}$, where $D_c$
is a diagonal matrix and V is some unitary matrix. Since
$tr(\vec{c}\cdot\vec{\sigma})^2=||\vec{c}||^2$, we have
$\lambda_{c}\leq ||\vec{c}||$, where$\lambda_{c}$ is the maximal
absolute value of eigenvalues of $\vec{c}\cdot\vec{\sigma}$. So
when (\ref{suff}) holds, $\mu_{\rm min}I+D_{c}\geq 0.$ Finally, we
have
\begin{equation}
\begin{array}{rl}
I\otimes \vec{c}\cdot\vec{\sigma}+UDU^{\dagger} &\geq
\mu_{\rm min}I+I\otimes \vec{c}\cdot\vec{\sigma}\\
&=
I\otimes V(\mu_{min}I+D_{c})V^{\dagger}\\
&\geq 0,
\end{array}
\end{equation}
where $V$ is unitary. This completes the proof.\qed

The term $I\otimes \vec{c}\cdot\vec{\sigma}+UDU^{\dagger}$ is
quite complicated, so one may hope to derive a simple expression
of $\vec{c}$, as what was done in the last section, which can be
used to characterize the positivity of the original term $I\otimes
\vec{c}\cdot\vec{\sigma}+UDU^{\dagger}$. Unfortunately, this is
impossible in general. The reason is that $UDU^{\dagger}$ and
$I\otimes \vec{c}\cdot\vec{\sigma}$ do not commute with each
other, but they are orthogonal to each other; in other words, we
have
$$tr(UDU^{\dagger}(I\otimes
\vec{c}\cdot\vec{\sigma}))=0$$ This implies that we cannot
diagonalize these two terms simultaneously. Otherwise, one of them
will be led into a trivial case. Suppose that
$$UDU^{\dagger}=I,\ \lambda_{p,q}=0,\ (p,q)\neq(0,0), \lambda_{0,0}=1.$$ To keep
$$I\otimes(I+\vec{c}\cdot\vec{\sigma})\geq 0,$$ it only needs
$$I+\vec{c}\cdot\vec{\sigma}\geq 0.$$ In other words, in this
special case, the only condition that guarantees the left-hand
side of the above inequality is positive is then not to move
$\vec{c}$ out the unit sphere (of course, we need to require that
$\vec{c}$ is a valid geometry representation of a quantum state).

\section{Conclusion}
In this paper, we use the set of generalized Pauli matrices as a
tool to present a new geometric representation of a quantum state
of qudits. This enables us to introduce affine mappings into the
space of qudits. Then we find a sufficient and necessary
condition for a special class of affine mappings.
This condition is also generalized to the case of $N$
qudits. Using this result, we can easily decide
whether a given mapping in the class we specified is a CP mapping
or not. As an application, we derive the optimal fidelity of the
universal $NOT$ on $d$-dimensional state space:
$$F(\rho^{\bot},\rho)=\frac{d}{d^2-1}.$$ Furthermore, we can easily deduce
that when $d=2$, or in the qubit case, this fidelity is optimal
with the value$$F(\rho^{\bot},\rho)=\frac{2}{3}.$$ The most
interesting thing that we discover here is that the CP condition
for this kind of affine mappings is closely related to the quantum
Fourier transform.

We also consider some more general affine mappings with diagonal
transition matrices and displacement vectors. We point out that
the problem of deciding the complete positivity of these mapping
is equivalent to the semi-positive definite programming problem.
So, it is impossible to derive a general symbolic solution of this
problem. However, we still get a
sufficient condition which says that when the norm of the
displacement vector is small enough, operation under consideration
is CP.

\smallskip\
\textbf{Acknowledgement:} We thank anonymous referee for helpful
comments and informing us of a valuable reference.

\end{document}